Quantum oscillations in the vortex state of underdoped YBa$_2$Cu$_3$O$_{6.5}$ and other multi-band superconductors

Lev P. Gor'kov*

National High Magnetic Field Laboratory, FL 32310, USA, and L. D. Landau Institute for Theoretical Physics, RAS, Chernogolovka 142432, Russia.

Abstract

We argue that the low- frequency quantum oscillations observed recently in the vortex state of underdoped *ortho* II-YBCO have the same origin as in other strongly correlated electronic systems. Superconductivity driven by strong interactions on several leading Fermi surfaces creates the proximity gaps on the other. The gap transferred on a small-sized pocket from larger Fermi surfaces is small. In case of unconventional pairing symmetry the induced gap is proportional to the ratio of oscillations' frequencies for small and large Fermi surfaces. The gap is in inverse proportion to the mass enhancement on the latter. We share the view that small pockets in themselves exist as a certain band feature among all Fermi surfaces.

(Dated: July 20, 2012)

Quantum oscillations (QO) in the mixed phase recently observed in the underdoped (UD) YBa$_2$Cu$_3$O$_{6,5}$ (the *ortho*-II YBCO) [1] are among the phenomena in high temperature superconductors (HTSC) that still remain poorly understood.

In metals the de Haas-van Alphen (dHvA) or the Shubnikov-de Haas (SdH) effects are powerful tools that, by measuring the frequencies of QO in the magnetic field at low temperatures, supply immediate information on the Fermi surfaces (FS). In view of high values of the upper critical fields, $H_{c2}$ and non-Fermi liquid character of the normal state, such experiments in HTSC cuprates deemed unpromising. Therefore, when the same methods revealed metallic FS for clean samples of UD YBa$_2$Cu$_3$O$_{6,5}$ in the vortex state [1], it was perceived with surprise, especially, because, considering the low frequency of oscillations, $F \approx 540$ T, the pocket's size is very small.

Below we argue that small Fermi surfaces, such as found in YBa$_2$Cu$_3$O$_{6.5}$ [1], are common to the whole group of strongly correlated superconductors (SC).

There is no consensus regarding the microscopic physics in cuprates. At the nominal oxygen concentration, YBa$_2$Cu$_3$O$_{6.5}$ the Fermiology would suggest that the CuO$_2$-planes in YBCO form metallic layers with the half-filled Brillouin zone (BZ). Such "large" FS had been seen only for strongly overdoped Tl$_2$Ba$_2$CuO$_{6+y}$ (Tl-2201) [2a, b]. In UD cuprates the Fermi liquid character of the electronic spectrum is destroyed by strong interactions; low-lying charge excitations exist only on "arcs" at the nodal points of "large" FS.

Several scenarios were proposed for origin of the pockets in cuprates. The most popular one asserts that pockets in YBCO arise from "large" FS restructured either at a phase transition, or by a "hidden order" in the pseudogap state above the superconducting dome. Any hypothetical incommensurate spin or density wave or the "stripe phase" that break translational symmetry may be the cause of "large" FS reconstruction. (See in [3a, b]).

We follow the interpretation of experiments [4]. Riggs et al [4] have studied the specific heat of the YBCO$_{6.56}$- crystal as the function of temperature in magnetic fields up to 45 T. Two specific heat components of the electronic origin were reliable identified in [4]. The one, featuring the square root dependence on the magnetic field, $\sqrt{H}$ is characteristic to the vortex phase of the $d$ –wave superconductivity; the second contribution that oscillates periodically as function of the inverse magnetic field, $1/H$ is due to normal carriers. The frequency of oscillations, $F \simeq 530\,\text{T}$ and the effective mass, $m^* \simeq 1.30\, m_e$ are in agreement with the results of other experiments for UD YBCO (for details, see in [4]).

Generally speaking, QO in the vortex phase prove only that the superconducting gap at the corresponding pocket is destroyed by magnetic fields. The key result [4] was that normal carriers persist down to the lowest temperatures (>1K) in the *zero fields*. With $F$ and $m^*$ already known independently, the authors could calculate the contribution into the electronic specific heat from the pocket and show it equal to the experimental value. Thereby, the analysis [4] firmly establishes the coexistence of the $d$ – wave SC and of the *single* pocket of "unpaired carriers" in the said range of parameters. QO were also seen in YBa$_2$Cu$_3$O$_8$ (YBCO124) (see in [3a,b]); "unpaired carriers" (H=0) were proven only for UD YBCO 123[4].

That several frequencies of the dHvA oscillations may survive in the vortex state well below the upper critical fields, $H<H_{c2}$ is known for other strongly correlated SC. Without a comprehensive understanding of HTSC, we concentrate first on those other SC where more information is available. We show that smallness of the pockets' sizes and symmetry of the order parameter are equally important for of the superconducting gap value on the pocket. Implications to UD YBCO are discussed later.

SC actively studied during last few decades all have the multi-band spectrum with properties often deviating from predictions of the weak-coupling microscopic theory. Non-gapped FS in the superconducting state below the upper critical field, $H_{c2}$, represent a new qualitative feature.

The first observation of the dHvA frequency in the mixed state dates back to the experiments by Graebner and Robbins [5] on 2$H$-NbSe$_2$ in 1976. The layered 2$H$-NbSe$_2$ is one of strongly correlated $s$-wave SC remarkable for high values of transition temperature, T$_c$ =7.2K and critical magnetic fields, $H_{c2\perp}$=3.2T and $H_{c2\parallel}$=13.2T (for the field directions perpendicular and parallel to the ($a$, $b$)-plane, correspondingly). The frequency, $F \approx 150\text{T}$ could be tracked down to $H_L$~2T, that is, deep into the vortex state: $H_L$ ~0.6$H_{c2}$; the effective mass is small, $m* = 0.61 m_e$. This frequency was assigned to the small three-dimensional Se-pocket. (Predicted in [6] were also the two large cylindrical FS at the $\Gamma$ - and the K-points of the BZ; the latter two are seen by ARPES [7], but have never been detected in the dHvA experiments).

The effect bears quite general character. It was observed in CeRu$_2$, URu$_2$Si$_2$ and UPd$_2$Al$_3$ (see review [8]); in the "high field" A15's [9]; in another heavy fermion (HF) CeCoIn$_5$ [10]; in borocarbides YNi$_2$B$_2$C and LuNi$_2$B$_2$C [11, 12]; in organic SC $\kappa$ -(BEDT-TTF)$_2$Cu(NCS)$_2$ [13] and $\beta^{"}$ -(BEDT-TTF)$_2$SF$_5$CH$_2$CF$_2$SO$_3$[14]. Notable, in all these examples the dHvA frequencies that persist in the mixed phase are small compared to the typical frequencies for FS in the material.

In the A15-superconductors, Nb$_3$Sn [9] and V$_3$Si [15] same frequencies are seen in the normal state and *also* in the mixed state ($F$~1500-2000T, $m^*$~1.7-2.7$m_0$). No higher frequencies have been detected and it remains unknown which FS is responsible for the superconducting transition.

Denote by $H_L$ the lowest field to which the frequency can be tracked down in the vortex state. $H_L$ varies from 0.2 $H_{c2}$ to 0.6 $H_{c2}$. QO in UD YBCO was first observed close to the irreversibility line ($\geq 25$T) [1]. In [4] QO in the specific heat were studied between 30T and 45T. $H_{c2}$ in UD YBCO may exceed 100T [16].

In late1990's the commonly accepted point of view was that at the field, $H$ close to the upper critical field, $|H-H_{c2}|<H_{c2}$, the gap remains small and electrons circle same quasi-classical orbits as in the normal phase. It was assumed as self-evident that the dHvA signal and the superconductivity itself belong to electrons on one and the same FS. Increased attenuation below $H_{c2}$ seemed natural, for electrons were expected to scatter on inhomogeneities both of the superconducting parameter and of the magnetic fields. (See in [17]). Meanwhile, the numerical solution for spectrum of a SC in the vortex state reveals oscillations *only* with the vortex lattice periodicity, $\propto B^{1/2}$ (see e.g., in [18]).

At the time it was not realized that the dHvA frequencies are so low. Small pockets with small masses contribute into superconductivity only weakly. Superconductivity sets in immediately below $H_{c2}$ due to pairing at other ("main") FS.

In [4] the normal carriers in the *zero fields* survive down to temperature ~1K, as if in YBCO$_{6.56}$ the pocket's gap, $\Delta_s$ were zero. The question arises whether the gap can be zero *identically*. Any small but finite gap must be destroyed when either the Landau quantization energy, $\hbar\omega_c = (e\hbar H/m^*c)$, or the temperature, T exceed $\Delta_s$. The question cannot be resolved in the dHvA-type experiments only, since observation of oscillations at lower fields is hindered by scattering of electrons on vortices.

Ideally, the normal carriers at low temperatures must manifest themselves in the specific heat experiments (as in [4]), in the thermal conductivity and in NMR. Smallness of the pocket makes such experiments difficult. Thus, the "selenium" pocket was not discerned in experiments [19] on the specific heat of 2H- NbSe$_2$.

Consider now which factors may be responsible for the especial smallness of the gap on the pocket. The electron-electron (*e-e*) interactions for any given band may differ in strength and even in signs. Not each FS contribute equally into the Cooper instability in multi-band metals. From relatively small effective masses (known from the dHvA data) one may conclude that electronic correlations on the pockets are weak and the pockets themselves possess no tendency to the Cooper pairing.

The gap on the pocket could be induced by the interband pairs' scattering. Within the weak-coupling scheme, the superconducting gap, $\Delta_i(\vec{k})$ on the FS (*i*) is:

$$\Delta_i(\vec{k}) = T\sum_{n,k} \int I_{i,k}(\vec{k},\vec{k}') F_k(i\omega_n;\vec{k}')[d\vec{k}'/(2\pi)^3]. \qquad (1)$$

$I_{i,k}(\vec{k},\vec{k}')$ are the matrix elements for interactions in the Cooper channel; FS are numbered by the indices (*i*, *k*). $F_k(i\omega_n;\vec{k}')$ is the anomalous Gor'kov function on the FS with index *k*:

$$F_k(i\omega_n;\vec{k}') = \Delta_k(\vec{k}')/\{\omega_n^2 + [\varepsilon_k(\vec{k}') - p_{Fk}^2/2m_k^{(b)}]^2 + |\Delta_k(\vec{k}')|^2\}. \qquad (2)$$

(Here $m_k^{(b)}$ is the *band* mass).

Integration over the energy and summation over the frequency, $\omega_n = (2n+1)\pi T$ in Eq. (1) lead to the familiar logarithmic factors:

$$\Delta_i(\vec{k}) = \ln(\bar{W}/T_c) \sum_k I_{i,k}(\vec{k},\vec{k}\,')\Delta_k(\vec{k}\,')v_k^{(b)}(E_F)d\Omega_{\vec{k}'}. \qquad (3)$$

Here, $\bar{W}$ is a cutoff, $d\Omega_{\vec{k}'}$ means integration along FS with index $k$, $v_k^{(b)}(E_F)$ is the bare (band) density of states (DOS) on the very same FS. With $T_c = \bar{W}\exp(-1/I_{eff}v_{av}(E_F))$ Eq. (3) acquires the form:

$$\Delta_i(\vec{k}) = \sum_k [I_{i,k}(\vec{k},\vec{k}\,')v_k(E_F)/I_{eff}v_{av}(E_F)]\Delta_k(\vec{k}\,')d\Omega_{\vec{k}'}. \qquad (4)$$

Eq. (4) establishes correspondence between gaps on different FS. The gap on a given pocket($s$), $\Delta_s$ is small, if the matrix elements that transfer pairing on the pocket are small compared to an effective pairing interaction, $I_{eff}$ on "main" FS: $\Delta_s \sim \Sigma_k[I_{s,k}v_k^{(b)}(E_F)/I_{eff}v_{av}(E_F)]\Delta_k$. (Here $v_{av}$ is the averaged DOS on "main" FS).

Not to crowd the equations, consider the case of only two FS (small($s$) and large ($l$)):

$$\Delta_s(\vec{k}) = \int [I_{s,l}(\vec{k},\vec{k}\,')/I_{eff}]\Delta_l(\vec{k}\,')d\Omega_{\vec{k}'}. \qquad (4a)$$

Matrix elements calculated on the wave functions belonging to bands of different origin usually have a numerical smallness: $(I_{i,k}/I_{eff}) < 1$. Then, even the gap in (4a) is small; it is finite and manifests itself at low enough temperatures.

Let us turn now to the momentum dependence in the matrix elements in Eq. (1). The general form is: $I_{i,k}(\vec{k},\vec{k}\,') = \sum_S I_{i,k}^S \chi_i^S(\vec{k}) \times \chi_k^S(\vec{k}\,')$ where summation runs over all symmetry representations, $S$ of the lattice space group. Single specific representation defines the superconducting symmetry.

Let the pocket be at a symmetry point of the BZ. Suppose that area occupied by the pocket in the BZ is small ($p_{F,s}a \ll 1$). Because interactions in the system vary on the lattice scale, $a$, in zero approximation one can neglect the dependence in the matrix element $I_{i,k}(\vec{k},\vec{k}\,')$ on the momentum $\vec{k}$. Therefore, in case of non-conventional symmetry, such as the $d$-wave symmetry, the transferred gap from Eq. (4a) is *zero*.

Non-gapped FS at low temperatures found so far only in CeCoIn$_5$ [20] and YBa$_2$Cu$_3$O$_{6.5}$ [4]. Notwithstanding all differences between cuprates and CeCoIn$_5$, in both cases superconductivity has the $d$-wave symmetry. Therefore, the above argument in favor of "zero gap" on the small pocket equally applies to UD YBCO. The experimental information, however, is richer for CeCoIn$_5$.

The induced $d$-wave gap cannot be zero *identically*. In Eq. (4a), consider now the dependence of $I_{s,l}(\vec{k},\vec{k}\,')$ on the momentum $\vec{k}$ in the next approximation. The vector $\vec{k}'$ lies on larger FS, $\vec{k} \ll \vec{k}'$; the first correction to $I_{s,l}(\vec{k},\vec{k}\,')$ proportional to the scalar product, $(\vec{k}\cdot\vec{k}')$ provides the angle-dependent factor, $kk'\cos\varphi$. In the second approximation such terms lead to the $d$-wave symmetry. By the order of magnitude the corresponding contributions are small as $|p_{F,s}a/\hbar|^2 \ll 1$. Area of the pocket and the oscillation frequency, $F$ are related: $p_F^2 = 2\pi\hbar^2(F/\phi_0)$. (Here, $\phi_0$ is the flux quantum, $\phi_0 \approx 2.10^{-7} Gauss*cm^2$). The factor $|p_{F,s}a/\hbar|^2$ proportional to the ratio, $F_s/F_L$ of small and "large" dHvA frequencies, after substitution into Eq.(4a) leads to:

$$\Delta_s^d \propto [I_{s,l}^d / I_l^d](F_s / F_L)\Delta_l^d . \qquad (5)$$

(The upper index, *d* in Eq. (5) denotes the *d*-wave pairing).

With $F_s \sim 0.6$ kT and $F_L \sim$ (5-12) kT in CeCoIn$_5$ [10] $F_s / F_L$ varies between 0.12 and 0.05. For cuprates ($F_s \sim 0.53$kT and $F_L \sim 18$kT [1, 2a, b]) this ratio is ~0.03. The STS-data [21a] for UD YBCO with Tc=60K gave $\Delta_l^d$ ~20meV; then $\Delta_s^d \sim (I_{s,l}^d / I_l^d)$ 6K. Experimentally, at smaller doping gaps become larger. From the thermal conductivity data [21b] $\Delta_l^d$ ~70meV for YBCO$_{6.54}$. With a smallness of $(I_{s,l}^d / I_l^d)$ in Eq. (5), both estimates by the order of magnitude do not contradict [4].

The factor $(I_{s,l}^d / I_l^d)$ can be small on purely numerical reasons, as mentioned before, but its value is determined mainly by interactions in the system. The discussion above was carried out in the weak-coupling scheme. Neither in CeCoIn$_5$, nor in YBCO interactions is weak.

With the notable exception of strong electron-phonon (*e-ph*) interactions in common metals [22], there is no comprehensive theory for strongly correlated systems. In the *e-ph* problem the Green function, $G^{-1}(i\omega_n;k) = i\omega_n - (\varepsilon(k) - \mu) - \Sigma(i\omega_n;k)$ near FS reduces to the following form:

$$G(i\omega_n;k) = Z / \{i\omega_n - Z[\varepsilon(k) - \mu]\}, \quad (6)$$

($\Sigma(i\omega_n;k)$ is the exact self-energy part due to interactions and $Z^{-1} = [1 - d(\Sigma(\omega;k = k_F) / d\omega)]$). The renormalized and the bare (band) electronic masses relate as $m^* = Z^{-1} m^{(b)}$ [22].

In the periodic Anderson model of strongly interacting fermions Eq.(6) for the Green functions appears in the mean-field approximation [23]. With $m^* = Z^{-1} m^{(b)}$ and the effective mass, $m^*$ known from the dHvA-type experiments, the $Z$-factor is thereby the measure of the interaction strength at *any given* FS. For the phonon-driven *s*-wave superconductors $Z^{-1}$ varies several-fold. In CeCoIn$_5$ $Z^{-1}$ is ~ 14 and ~28 for two frequencies, $\beta_1, \beta_2$, correspondingly [10].

The mean-field Green functions Eq. (6) slightly changes the anomalous function, $F_k(i\omega_n;\vec{k}')$ in Eq. (2). The self-consistency equation (one FS, for simplicity) that determines temperature of the superconducting transition, T$_c$ is given by Eqs. (1,2) linear in $\Delta(\vec{k})$:

$$\Delta(\vec{k}) = T \sum_n \int |I_l(\vec{k};\vec{k}')| \frac{Z^2 \Delta(\vec{k}')[d\vec{k}' / (2\pi)^D]}{(\omega_n)^2 + [Z(\varepsilon(k') - \mu)]^2} . \qquad (7)$$

(D=2 or 3, stands for dimensionality). After the frequency summation and the integration in the momentum space, the criterion (7) is practically of the same form of Eq. (3), except for the $Z$-factor in $Zv^{(b)}(\varepsilon_F)$:

$$\Delta(\vec{k}) = \ln(\bar{W} / Tc) \int |I_l(\vec{k};\vec{k}')| Zv^{(b)}(\varepsilon_F) \Delta(\vec{k}') d\Omega_{\vec{k}'} . \qquad (8)$$

From Eq. (8) it follows that $\ln(\bar{W}/T_c) \sim [Z|I_l|\nu^{(b)}(\varepsilon_F)]^{-1}$. In the HF physics $\bar{W}$ is the Kondo temperature, $T_K \sim$ (15-30) K. The bare (the band) DOS, $\nu^{(b)}(\varepsilon_F)$ has its usual value of several states per 1eV. Hence, small $Z$ in Eq.(8) must be compensated by an interactions' enhancement(see, e.g., [24]). Then $I_l(\vec{k};\vec{k}') \propto Z^{-1}$ and Eq. (5) acquires an additional small factor:

$$\Delta_s^d \propto (m^{(b)}/m^*)(F_s/F_L)\Delta_l^d \quad . \quad (9)$$

The thermal conductivity of CeCoIn$_5$ was recently measured for temperatures down to ~50mK [25]. Contrary to the previous finding of the *zero* gap on one FS [20], small but finite *d*-wave gap, $\Delta_s^d \sim$ 0.01K was found in [24]. Its value is consistent with Eq.(9) where the product $(m^{(b)}/m^*)(F_s/F_L)$ is between $\sim 10^{-3}-10^{-2}$ (T$_c$=2.3K).

The periodic Anderson model is not applicable to YBCO. Interactions are strong, but the Cooper instability in cuprates must be treated differently. Independently of its specific mechanism the pairing couples carriers on the opposite parts of "large" FS. The logarithmic singularity $\ln(\bar{W}/T)$ comes about only from low-lying electronic excitations *inside* the "arc" so that interactions must be strong to compensate narrowness of "arcs". These arguments indicate in the same direction as for CeCoIn$_5$.

Beside UD YBCO, QO with the small frequency (F~300T) were reported in the electron-doped Nd$_{2-x}$Ce$_x$CuO$_4$ (NCCO) [26, 27]. The interpretation was based on a picture of reconstructed "large FS" deformed along the (0, 0)-$(\pi,\pi)$ direction and is similar to the one discussed for QO in YBCO [1,3]. In particular, it was argued that "large" FS is seen only via "magnetic breakdown" (MB) effects [26]. The would-be MB trajectory runs along postulated the electron- and the hole-like pockets positioned at $(\pi,0)$ and $(\pi/2,\pi/2)$, correspondingly. The applicability of such picture to the electron-doped NCCO is again undermined by the very fact that only single pocket at $(\pi,0)$ was seen in [26, 27]. Moreover, only the electronic pockets at $(\pi,0)$ with the area proportional to x were independently detected by ARPES [28]. (The frequency of QO gradually decreases with doping [26]; this can be ascribed to changes in the band structure with doping). Still, the situation with QO in the electron-doped NCCO needs further clarification.

Unlike NCCO, ARPES [29] could not discern small pockets in UD YBCO because quality of the surface is hindered by the CuO-chain layers. (The estimate for the antinodal gap ~80meV [29] was consistent with the data [21b]).

To conclude, we argue that not all FS play equal role in superconductivity in multi-band SC. Large FS with stronger *e-e* interactions contribute most into the Cooper instability. Strong interactions on such FS manifest themselves in enhanced effective masses. Study of QO from such point of view should help at interpretation for pairing mechanisms in a number of poorly understood materials.

The interband matrix elements inevitably transfer the superconducting order from "main" FS to the rest. We demonstrated that transferred gaps can be small. Apart of a natural smallness of matrix elements between different zones, strong interactions on "main" FS diminish the transferred gap by the factor $\propto m^{(b)}/m^*$. Small gaps can be destroyed in the magnetic field, thus, resulting in observation of QO in the vortex state.

Strong interactions enlarge electronic contribution to the specific heat. Many SC reveal large Sommerfeld coefficients, however, there was no attempts yet to correlate it with the observation of QO below $H_{c2}$. An intriguing example is presented by QO in some organic conductors [13, 14a, b]. Theoretically, their spectrum consists of the two open sheets and one closed FS. However, in $\beta''$-(BEDT-TTF)$_2$ SF$_5$CH$_2$CF$_2$SO$_3$ such FS is much smaller than expected(~5% of the BZ). QO have the pronounced 2D character with the chemical potential probably *fixed* due to large DOS at open sheets. This indicates on strong e-e interactions there [14b].

Non-conventional symmetry drastically diminishes the gap if the pocket lies on the symmetry planes where the d-wave changes sign. Then the transferred gap of Eq. (9) may stay "invisible" down to low temperatures. We suggested this explanation for observation of "normal" carriers in [4] down to ~1K. The compelling argument to confirm such interpretation would be detection of a finite gap at temperatures below 1K. Such feature was verified for CeCoIn$_5$. For *ortho*-II YBCO the specific heat experiments further below 1K are difficult. We suggest instead the thermal conductivity experiments, as in case of CeCoIn$_5$ [20, 25], or NMR. Note that applicability of Eqs.(5,9) does not depend on the origin of pockets.

For the whole group of multi-band SC small pockets come as a detail of the electronic band structure; the frequencies are usually identifiable with proper FS (several pockets are possible). For UD YBCO123 the conjecture was expressed first in [1]. Band calculations [30, 31] do not contradict this scenario in YBCO123, but not in YBCO124 [30]. Still, only single QO frequency was seen so far in the two-chain YBCO [32].

The author is grateful to O. Vafek for numerous discussions concerning experimental details and the interpretation of data [4]. The author thanks M. Tanatar for bringing to his attention Ref. 23.

The work was supported by the NSF Cooperative Agreement No. DMR-0654118 and by the State of Florida.

* Electronic address: gorkov@magnet.fsu.edu